\documentclass[letter]{jpsj3}
\usepackage{txfonts}
\usepackage{graphicx}
\usepackage{ulem}
\title{
Induced discommensurations in the lock-in transition of charge-density waves
}
\author{
Katsuhiko Inagaki$^1$\thanks{kina@asahikawa-med.ac.jp} and Satoshi Tanda$^2$
}

\inst{$^1$ Asahikawa Medical University, Midorigaoka Higashi 2-Jo, Higashi 1-Chome, Asahikawa, 078-8510, Japan\\
$^2$ Division of Applied Physics, Graduate School of Engineering, Hokkaido University, Kita 13 Nishi 8, Kita-ku, Sapporo 060-8628, Japan}

\abst{
We studied McMillan's free energy of the lock-in transition in charge-density waves. The wave profiles near the critical value were obtained by numerical annealing. First, we demonstrated that our method reproduces the previous studies. The obtained wave profiles include discommensurations near the critical value. Then, we calculated possible wave profiles in the commensurate state.  
We found that discommensurations are able to be excited in the commensurate state, leading the system to turn into an incommensurate state. We proposed that these wave profiles result from topological invariants. 
Moreover, excitation of the discommensurations is favorable for the direction to the original wavelength of the incommensurate state. This is attributed to the nature of McMillan's free energy. The current-induced incommensurations, which we discovered with the diffraction study of $o$-TaS$_3$ [Inagaki et al., J. Phys. Soc. Jpn. 77, 093708 (2008)], is consistent with this study.
}

\begin{document}
\maketitle

The lock-in transition in charge-density waves (CDWs) has been an intriguing issue due to the interaction between two length scales: the Fermi wavelength and a lattice constant. Near the transition between commensurate and incommensurate CDWs, periodic phase solitons, or discommensurations, occur. McMillan first predicted this phenomenon; then, Bak and Emery obtained a similar result using a different approach \cite{Bak1976, McMillan1976}. This model has been well studied for quasi-two-dimensional systems \cite{Nakanishi1977, Nakatsugawa2024}. However, the theories have not fully explained experimental results, particularly of quasi-one-dimensional conductors \cite{Sagar2008, Inagaki2008}, such as $o$-TaS$_3$ and K$_{0.3}$MoO$_3$. 
In this article, we performed a numerical study of McMillan's free energy, which describes the lock-in transition. We found topologically protected approximate solutions in the commensurate state. These solutions include discommensurations, hence the system turns out to be incommensurate. These solutions may explain why the flowing current changes the intensities of diffraction peaks of $o$-TaS$_3$, proportional to volume fractions, from commensurate to incommensurate CDWs in the coexistence regime\cite{Inagaki2008}. This is also consistent with the `soliton transportation', by which anisotropy in conduction is observed to be increased  \cite{Takoshima1980, Latyshev1983, Rojo-Bravo2011}.  

We performed numerical optimization of McMillan's free energy \cite{McMillan1976, Inagaki2024}
\begin{eqnarray}
f&=&\int dx [Y(1-\cos 3\theta ) +(d \theta /dx -1)^2],\label{eq:energy}
\end{eqnarray}
where $f$ is the reduced energy, $Y$ is a parameter corresponding to the temperature, and $\theta$ is the phase at the position $x$. 
This formula was derived from the Ginzburg-Landau equation for the CDW order parameter \cite{McMillan1976}. At the Peierls temperature, the gap begins to develop with incommensurate CDWs. The lock-in transition is determined by the phase $\theta$.
If $\theta(x)=0$, the system is commensurate, while if $\theta(x)=x$, the system is purely incommensurate. The contribution of the third-order Umklapp process is represented by a temperature-dependent parameter $Y$. At the Peierls temperature, $Y=0$, while if $Y=Y_c$, the lock-in transition undergoes. Previous studies \cite{Fukuyama1985} revealed the critical value $Y_c=\pi^2/8=1.2337 \cdots$.
Although the analytic solution was already given, there remain some phenomena beyond the solution. Hence, we explored other possibilities by simulated annealing, which might provide approximate solutions from given initial conditions. This method was first proposed by Metropolis and coworkers to find approximate solutions for optimization problems \cite{Metropolis1953}. 

In this study, we followed the standard procedure of simulated annealing by introducing the dimensionless temperature $T_a$. This parameter determines the Boltzmann distribution at $T_a$ for the wave profile.  The initial conditions were pure incommensurate phase, $\theta=sx$, where $s$ is an initial slope ($0<s<1$).
A trial randomness $\varepsilon$ was introduced into $d\theta/dx$ at the site $i$, namely $d\theta/dx + \varepsilon$, and the phases $\theta$ for $j>i$ were shifted by $\varepsilon$. This process is an analogue of the phase modulation in CDWs. With respect to the boundary condition, $\theta(0)=0$, which was the only constraint in the numerical calculations. The phase at the other end can be varied. The new wave profile was accepted or discarded according to the probability $e^{-f'/T_a}$, where $f'$ is the free energy (\ref{eq:energy}) of the new wave profile.
By iteration of introducing such a randomness, the wave profile deformed to be in equilibrium at $T_a$. Then the parameter $T_a$ was decreased, and the wave profile was optimized. Finally, we set $T_a=0$ to obtain the wave profile with the lowest energy.

\begin{figure}
\vspace*{6em}
\includegraphics[width=0.8\textwidth]{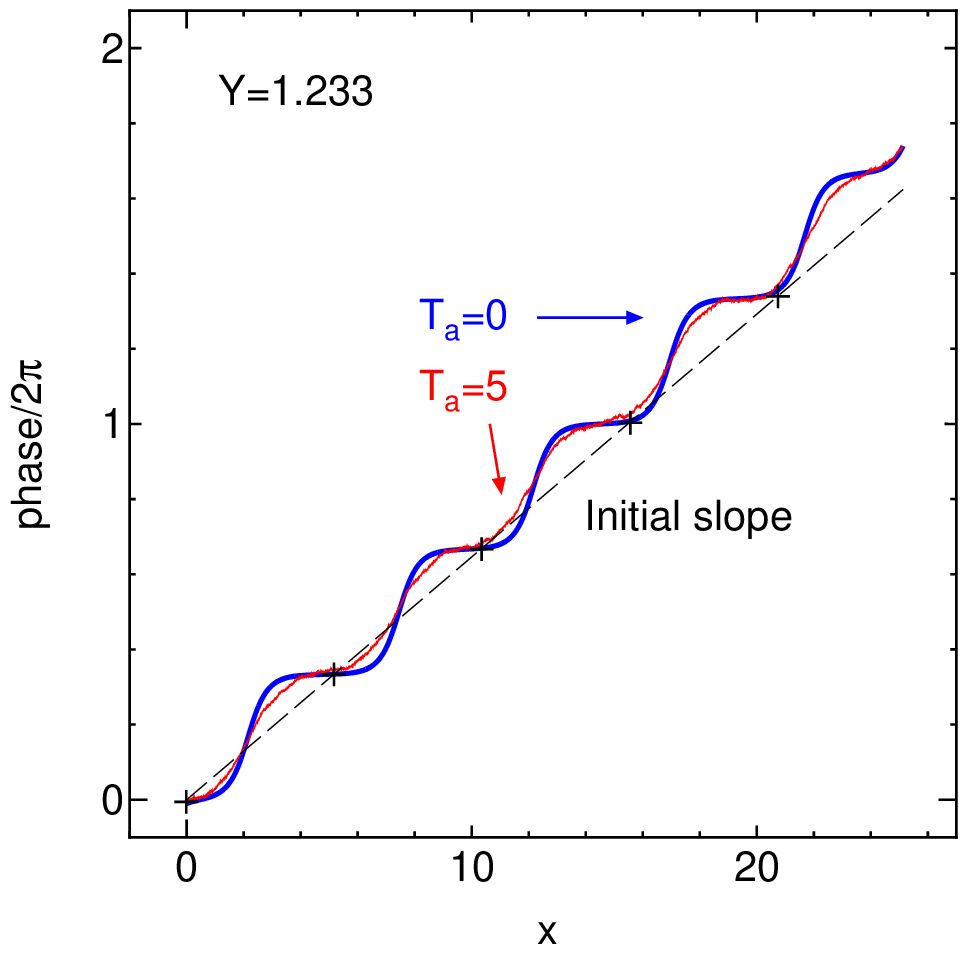}
\vspace*{1em}
\caption{(Color online) Optimization of the wave form from a pure incommensurate CDW.
The broken line represents the initial phase. The thin and thick solid lines show the intermediate ($T_a=5$) and final phases, respectively. The crosses represent the initial phase at $2n\pi/3$ ($n$ is an integer).}\label{fig:optimization}
\end{figure}

Figure \ref{fig:optimization} shows the initial (broken line), the intermediate (thin solid line), and the final (thick solid line) wave profiles in optimization at $Y=1.233$. As the temperature $T_a$ rises, the wave profile changes to the equilibrium state. During this first procedure, the overall shape of the wave profile is formed, as shown in the thin solid line. Then the temperature $T_a$ gradually decreases to zero.  The wave profile becomes smooth to release the energy accompanied with the $d\theta/dx$.

\begin{figure}
\vspace*{5em}
\includegraphics[width=0.8\textwidth]{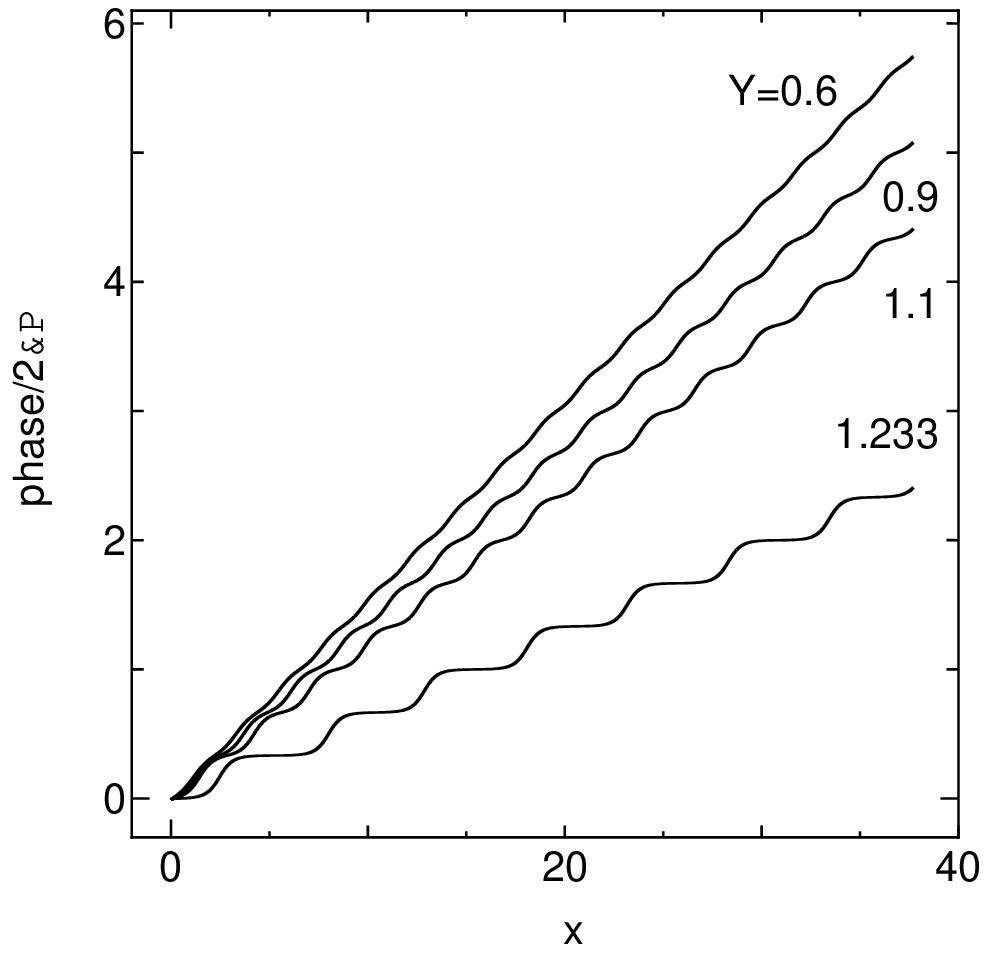}
\vspace*{1em}
\caption{Evolution of wave profiles. The parameters $Y=0.6$, 0.9, 1.1, and 1.233 from top to bottom. Discommensurations emerge significantly near the critical value, accompanying $2\pi/3$ phase jumps. The wave form at $Y=1.233$ shows that our study reproduces McMillan's first study on discommensurations \cite{McMillan1976}.}\label{fig:evolution}
\end{figure}

We first checked the reproducibility of the known results. Figure \ref{fig:evolution} shows the wave profiles for $Y=0.6$, 0.9, 1.1, and 1.233. When $Y$ approaches the critical value $Y_c$, the average slope of wave profiles gets closer to commensuration, namely, $\theta=0$.  We confirmed that the free energies agree with those obtained in the previous study within numerical errors. The emergence of discommensurations was also reproduced. It should be noted that in the previous studies, the superposition of a linear slope and a Fourier series was assumed as an ansatz, and calculations were performed over its period \cite{McMillan1976, Inagaki2024}. On the other hand, the periodicity in the obtained wave profiles emerged spontaneously in this study. This shows an advantage of optimization by simulated annealing. As we mentioned in the previous study \cite{Inagaki2024}, McMillan's free energy Eq. (1) does not include interaction between discommensurations. Although the periodicity is not rigorous, the occurrence of the discommensurations, or a train of solitons, does not need the interaction, as shown in Figs. \ref{fig:optimization} and \ref{fig:evolution}. 

\begin{figure}
\vspace*{5em}
\includegraphics[width=0.8\textwidth]{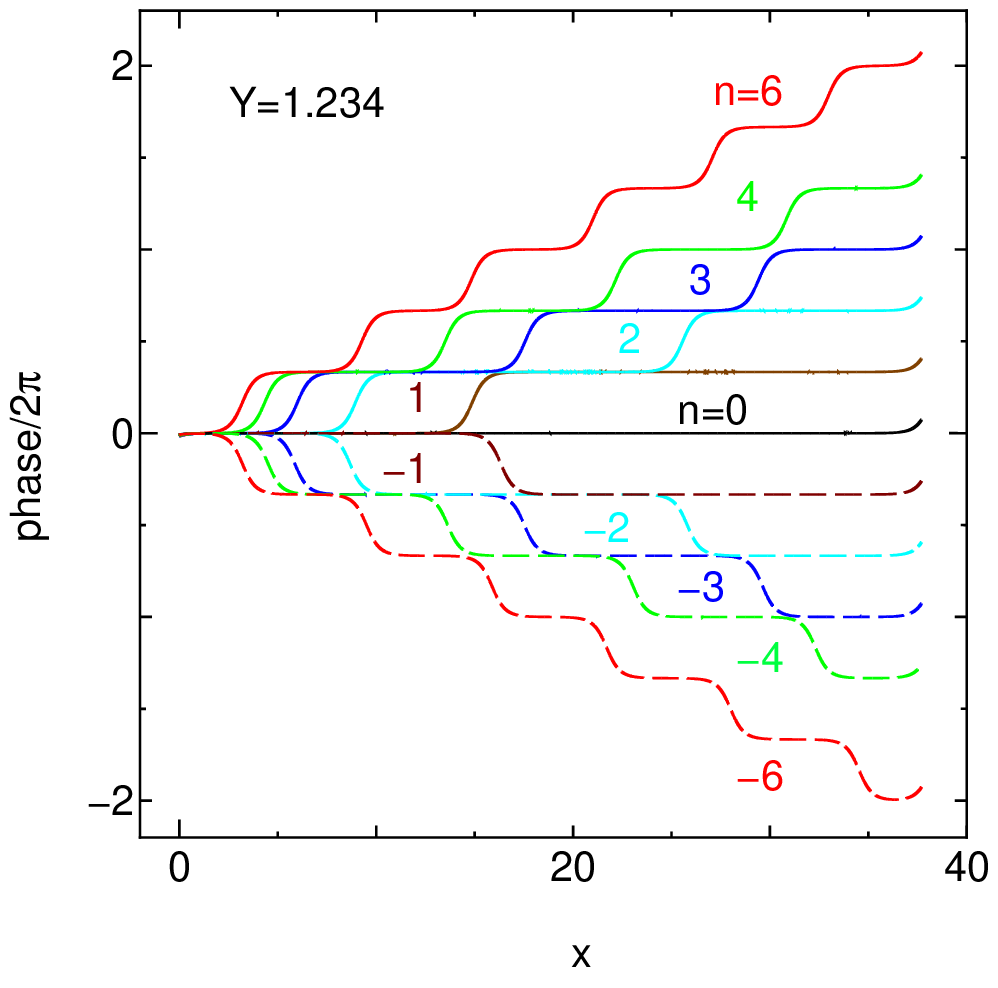}
\vspace*{1em}
\caption{(Color online) Wave profiles at $Y=1.234$, slightly higher than $Y_c$. By changing the initial slope, the obtained wave profiles include discommensurations for $n \ge0$ (solid lines) and $n<0$ (broken lines).}\label{fig:Yc}
\end{figure}

Figure \ref{fig:Yc} shows the wave profiles at $Y=1.234$, slightly higher than $Y_c$. To include discommensurations, the initial average slopes were set as $2\pi n/120$ ($n=0$, $\pm1$, $\pm2$, $\pm3$, $\pm4$, and $\pm6$), then numerical annealing was performed for each $n$. The solid and broken lines are the results for $n \ge 0$ and $n <0$, respectively.
As shown in Fig. \ref{fig:Yc}, each discommensuration has the phase difference of $\pm 2\pi/3$. Since $Y>Y_c$, these wave profiles cannot be represented by the known analytic solution of Eq. (1). The behavior of Eq. (1) is understood as motions of a pendulum. 
For $Y<Y_c$, the pendulum makes circular motions, corresponding to discommensurations. On the other hand, the pendulum only oscillates near the origin for $Y>Y_c$, where discommensurations cannot be created in the wave profiles. Hence, the interpretation of the obtained wave profiles requires deep insight.

We focus on the topological nature in the systems described by Eq. (1). The first term of the integrand in Eq. (1) is equivalent between phase $\theta$ and $\theta+2\pi/3$. In particular, $\cos 3\theta= \cos 2\pi = 1$ for $\theta=2 m\pi/3$, where $m$ is an integer. At each point $x$ with $\theta=2m\pi/3$, the local energy around $x$ should be minimum due to Eq. (1). Each discommensuration only occurs by deforming the phase path between the two points with $m$ and $m+1$. Hence,
 the number of intersections of the phase profile with $2m\pi/3$ must be a topological invariant even if the wave profile changes, as shown in Fig. \ref{fig:optimization}.This explains the reason why
the resulting wave profiles at $Y=1.234$ include discommensurations (Fig. \ref{fig:Yc}).
As we described above, we started from the pure incommensurate states. The number of the intersections was determined by the initial conditions, and protected by topology.

\begin{figure}
\vspace*{6em}
\includegraphics[width=0.8\textwidth]{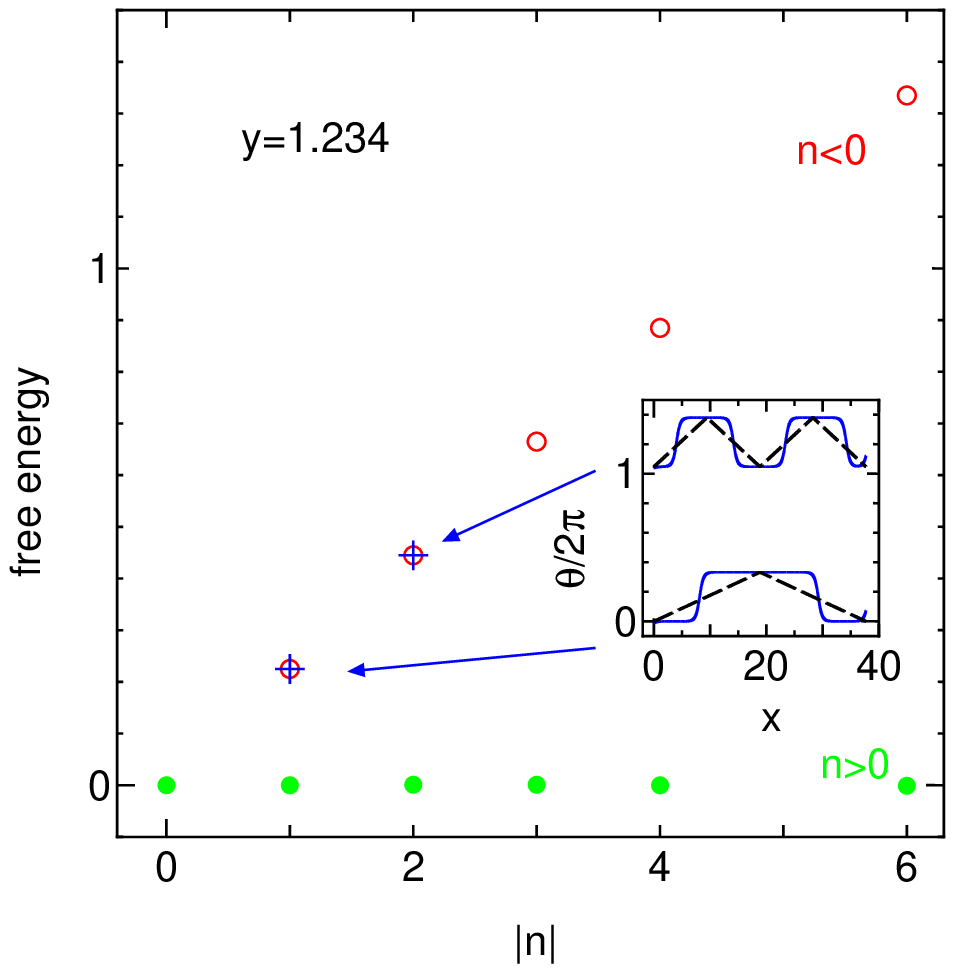}
\vspace*{1em}
\caption{(Color online) Free energies for discommensurations for $n>0$ (open circles) and $n<0$ (solid circles). By increasing the number of discommensurations, the energy does not increase. On the contrary, the energy is proportional to the number of anti-discommensurations. For comparison, the free energies for soliton-antisoliton pairs are also plotted (crosses). The energies of the pairs coincide with those of discommensurations for $n<0$. Inset: Wave profiles used in the calculation of the soliton-antisoliton energies. The broken and solid lines show the initial and final phases, respectively. The top curves with two pairs are shifted by 1. The average slope of both curves is zero, suggesting that the extra energy is localized at antisolitons.}\label{fig:energy}
\end{figure}

One may ask if discommensurations are equivalent for $n>0$ and $n<0$ because the obtained wave profiles look symmetrical against the sign of $n$. We calculate the extra energy relative to the commensurate state ($n=0$) for these wave forms. Figure \ref{fig:energy} shows the result. The free energy is plotted for each number of discommensurations (solid and open circles for positive and negative $n$'s, respectively). Interestingly, for $n>0$, the free energy of discommensurations seems to be zero within numerical errors. Lock-in transition described in Eq. (1) was found to be second-order \cite{McMillan1976, Inagaki2024}. Since the calculations are performed at slightly higher than the critical value, the obtained wave profiles with discommensurations must be induced in incommensurate metastable states. In contrast, the free energy is proportional to the number of discommensurations for $n<0$. This is natural because the second term of the integrand in Eq. (1) provides extra energy from the initial slope being 1, at which the Peierls transition occurs. Therefore, commensurations are not symmetrical in terms of energy.

This result leads to the next question: for $n<0$, can the extra energy in discommensurations be attributed to either the overall negative slope of the system or to localized discommensurations? 
Hence, we investigated the wave profiles, including soliton-antisoliton pairs. The inset of Fig. \ref{fig:energy} shows the obtained wave profiles (solid lines), starting from triangular waves (broken lines). As we described above, the number of intersections of $2\pi/3$ should be maintained during optimization processes. Since we started with triangle waves, soliton-antisoliton pairs emerged as expected. The difference from the previous calculation is that the overall slope is zero, even though the wave profiles include solitons and antisolitons. The extra energy for each case is plotted as the crosses in Fig. \ref{fig:energy}. It is found that the energy coincides with that of the case for $n<0$, suggesting that the energy is localized at antisolitons.

This leads to a new insight that soliton-antisoliton pairs are unlikely to be excited in McMillan's framework. We have found that discommensurations are energetically unfavorable for $n<0$, as shown in Fig. \ref{fig:energy}, and the energy is localized at each one. This differs from soliton-antisoliton excitation, discussed by Maki, followed by Duan and Hatakenaka, all of whom studied soliton-antisoliton pairs in CDWs \cite{Maki1977, Duan1993, Hatakenaka1998}. In their studies, a soliton-antisoliton pair is assumed to be excited by quantum tunnelling. Both soliton and antisoliton have extra boundary energies, while the energy lowers between the pair. By considering them, the height of tunnel barriers was calculated. 
It should be noted that the interaction between (anti-)solitons and the electric fields created by the solitons is not taken into account in this study. This explains the difference between our finding and the previous studies.

We compared our findings with the experimental result of the lock-in transition in $o$-TaS$_3$. This quasi-one-dimensional conductor undergoes the Peierls transition at 218 K to appear incommensurate CDWs. At lower temperatures, the wavelength of the incommensurate CDWs shifts toward a commensurate one; however, commensurate and incommensurate CDWs coexist for 50-130 K \cite{Inagaki2008}.
In the coexistence region, the intensity of incommensurate CDWs was reported to be increased by flowing current, while that of commensurate CDWs was decreased. This phenomenon was interpreted as current-induced incommensurate CDWs. The present study is consistent with the experiment where commensurate CDWs tend to return to the incommensurate state, as shown in Fig. \ref{fig:energy}. This implies that incommensurate CDWs are easier for current to flow through than commensurate ones. 
We should also be reminded that the system is quasi-one-dimensional. Induced incommensurate CDWs carry electric current along the chain direction. In other words, anisotropy in electric conduction becomes large. This scenario is what we have known as the `soliton transport', observed in $o$-TaS$_3$ in early studies of conductivity and Hall resistance \cite{Takoshima1980, Latyshev1983}. Since our study is based on McMillan's free energy Eq. (1), the transition is second-order \cite{McMillan1976, Inagaki2024}. Hence it should be careful to apply our result to the experimental facts. 
At least, both of the current-induced incommensurate CDWs and the soliton transport phenomena in $o$-TaS$_3$ are consistent with our finding. 

The fractional charge accompanied with discommensurations have been predicted.
The ratio between the CDWs wavelength $\lambda$ and a lattice constant $a$ is an integer $M=\lambda/a$. Integration of the CDW phase over the $2\pi$ period emerges the charge $2e$. In the discommensurate phase, there are $M$ solitons in the period, suggesting that the fractional charge $2e/M$ is accompanied with each soliton. 
The diffraction study of K$_{0.3}$MO$_3$ revealed evidence of discommenusation \cite{Rojo-Bravo2011}. The wave profiles were reconstructed by the diffraction patterns. Interestingly, the phase jump at soliton was $2\pi$ (or charge $2e$), instead of the expected value of $\pi/2$ (commensuration ratio $M$=4). The Aharonov-Bohm oscillation observed in the ring crystal of $o$-TaS$_3$ also suggested the charge $2e$ \cite{Tsubota2012}.
This discrepancy remains unsolved within this study. 

Finally, a recent study suggests that the Umklapp term induces first-order transitions between incommensurate and commensurate CDWs, and in more general cases \cite{Rozhkov2025}. This study also provides possible metastable states in incommensurate CDWs. However, discommensurations in CDWs are out of focus of the study. There is an unrevealed nature behind the lock-in transition of density waves. 

\section*{Acknowledgements}
We appreciate K. Nakatsugawa and S. Kashimoto for fruitful discussions.

\begin{thebibliography}{99}
\bibitem{Bak1976}
P. Bak and V. J. Emery, Phys. Rev. Lett. 36, 978 (1976).
\bibitem{McMillan1976}
W. McMillan, Phys. Rev. B 14, 1496 (1976).
\bibitem{Nakanishi1977}
K. Nakanishi and H. Shiba, J. Phys. Soc. Jpn. 43, 1839 (1977).
\bibitem{Nakatsugawa2024}
K. Nakatsugawa, T.N. Ikeda, T. Toshima, and S.Tanda,
Phys. Rev. \textbf{B 109}, L081407 (2024).
\bibitem{Sagar2008}
D. M. Sagar, D. Fausti, S. Yue, C. A. Kuntscher, S. van Smaalen, and
P. H. M. van Loosdrecht, New J. Phys. 10, 023043 (2008).
\bibitem{Inagaki2008}
K. Inagaki, M. Tsubota, K. Higashiyama, K. Ichimura, S. Tanda, K. Yamamoto, N. Hanasaki, N. Ikeda, Y. Nogami, T. Ito, and H. Toyokawa,
J. Phys. Soc. Jpn. 77, 093708 (2008).
\bibitem{Takoshima1980}
T. Takoshima, M. Ido, K. Tsutsumi, T. Sambongi, S. Honma, K. Yamaya, and Y.Abe,
Solid State Commun. 35, 911 (1980).
\bibitem{Latyshev1983}
Yu.I. Latyshev, Ya.S. Savitskaya, and V.V. Frolov,
JETP Lett. 38, 541 (1983).
\bibitem{Rojo-Bravo2011}
A. Rojo-Bravo, V.L.R. Jacques, and D. Le Bolloc'h,
Phys. Rev. B 94, 201120(R), (2016).
\bibitem{Inagaki2024}
K. Inagaki, K. Nakatsugawa, and S. Tanda,
J. Phys. Soc. Jpn. 93, 083702 (2024).
\bibitem{Fukuyama1985}
H. Fukuyama and H. Takayama, in Electronic Properties of Inorganic
Quasi-One-Dimensional Materials, I, ed. P. Monceau (Reidel,
Dordrecht, 1985) p. 41.
\bibitem{Metropolis1953}
Nicholas Metropolis, Arianna W. Rosenbluth, Marshall N. Rosenbluth, Augusta H. Teller, Edward Teller,
J. Chem. Phys. 21, 1087-1092 (1953).
\bibitem{Maki1977}
K. Maki, Phys. Rev. Lett. 39, 46 (1977).
\bibitem{Duan1993}
J.-M. Duan, Phys. Rev. B 48, 4860 (1993).
\bibitem{Hatakenaka1998}
N. Hatakenaka, M. Shiobara, K. Matsuda, and S. Tanda,
Phys. Rev. B 57, R2003 (1998). 
\bibitem{Tsubota2012}
M. Tsubota, K. Inagaki, M. Matsuura, and S. Tanda,
Europhys. Lett. 97, 57011 (2012).
\bibitem{Rozhkov2025}
A.V. Rozhkov, Phys. Rev. B 111, 205434 (2025).
\end{thebibliography}
\end{document}